\documentclass[prl,twocolumn,showpacs]{revtex4}
\usepackage[dvips]{graphicx}
\usepackage{latexsym}
\usepackage{bm}
\begin{document}
\newcommand{\fig}[2]{\includegraphics[width=#1]{#2}}
\newcommand{\pprl}{Phys. Rev. Lett. \ }
\newcommand{\pprb}{Phys. Rev. {B}}
\newcommand{\be}{\begin{equation}}
\newcommand{\ee}{\end{equation}}
\newcommand{\bea}{\begin{eqnarray}}
\newcommand{\eea}{\end{eqnarray}}
\newcommand{\nn}{\nonumber}
\newcommand{\la}{\langle}
\newcommand{\ra}{\rangle}
\newcommand{\dg}{\dagger}
\newcommand{\upa}{\uparrow}
\newcommand{\dna}{\downarrow}

\title{Electron correlation and Fermi surface topology of Na$_x$CoO$_2$}

\author{Sen Zhou$^1$, Meng Gao$^1$, Hong Ding$^1$, Patrick A. Lee$^2$,
and Ziqiang Wang$^1$}
\affiliation{$^1$ Department of Physics, Boston College, Chestnut Hill,
MA 02467}
\affiliation{$^2$ Department of Physics, Massachusetts Institute of Technology,
Cambridge, MA 02139}

\date{\today}

\begin{abstract}

The electronic structure of Na$_x$CoO$_2$ revealed by recent photoemission
experiments shows important deviations from band theory predictions.
The six small Fermi surface pockets predicted by LDA calculations have not
been observed as the associated $e_g^\prime$ band fails to cross the
Fermi level for a wide range of sodium doping concentration $x$. In addition,
significant bandwidth renormalizations of the $t_{2g}$ complex
have been observed. We show that these discrepancies are due to strong
electronic correlations by studying the multi-orbital Hubbard model
in the Hartree-Fock and strong-coupling Gutzwiller approximation.
The quasiparticle dispersion and the Fermi surface topology obtained
in the presence of strong local Coulomb repulsion are in good agreement
with experiments.

\typeout{polish abstract}
\end{abstract}

\pacs{71.27.+a, 71.18.+y, 74.25.Jb, 74.70.-b}

\maketitle
The cobaltates (Na$_x$CoO$_2$) are doped 3d transition metal oxides in which
the Co atoms form a layered hexagonal lattice structure. In contrast to
the high-$T_c$ cuprates, where the Cu$^{2+}$ has a 3d$^9$ configuration and
occupies the highest {\it single} $e_g$ ($d_{x^2-y^2}$) orbital near the
Fermi level, the cobaltates are {\it multi-orbital} systems where
the Co$^{4+}$ is in the 3d$^5$ configuration, occupying the three
lower $t_{2g}$ orbitals, similar to the ruthenates (Sr$_2$RuO$_4$).
The unexpected discovery \cite{takada03} of a superconducting phase of yet
unknown origin in hydrated Na$_x$CoO$_2$ around $x\sim0.3$ has generated
renewed interests in this material. However, such basic issues as the
low energy electronic structure and Fermi surface topology in the
cobaltates have not been well understood.
Band structure (LDA) calculations \cite{singh00}
find that the trigonal symmetry of the Co site in the triangular
lattice splits the three $t_{2g}$ complex into an $a_{1g}$
and two degenerate $e_g^\prime$ states at the zone center ($\Gamma$ point).
The LDA predicts a large Fermi surface (FS) associated with the $a_{1g}$
band enclosing the $\Gamma$ point and
six small FS pockets of mostly $e_g^\prime$ character near the K points
\cite{singh00,pickett}.

However, recent angle-resolved photoelectron spectroscopy (ARPES)
measurements on the cobaltates revealed
only a {\it single} hole-like FS centered around the $\Gamma$ point for
a wide range of Na concentration $x$ \cite{hbyang2,hasan2,hasan,hbyang1}.
The area enclosed
by the FS exhausts the Luttinger volume, which is consistent with the
observation that the dispersion of the $e_g^\prime$ band associated
with the LDA FS pockets lies below and
never crosses the Fermi level \cite{hbyang2}.
The absence of the FS pockets is unexpected and puts serious
constraints on several proposed theories of non-phonon mediated
superconductivity as well as magnetic properties based on the
nesting conditions of the FS pockets \cite{kuroki04,johannes04,yanase}.
Furthermore, the measured quasiparticle bandwidths are significantly
narrower than the LDA predictions \cite{hbyang2,hasan2}.
These fundamental discrepancies between ARPES and LDA suggest that the
effects of strong electronic correlations are important in the cobaltates.
The effects of local Coulomb repulsion $U$ has been considered
in the LSDA+U approach, which indeed finds the absence of the
small FS pockets \cite{PHZhang}. However, the latter is tied to
the fully polarized ferromagnetic state in the LSDA+U theory which
gives spin-split bands and a spin polarized FS with an area twice as large.
This is inconsistent with ARPES
and likely an artifact of the LSDA+U approximation.
A recent calculation based on the multi-orbital Hubbard model
and the dynamical mean-field theory finds that the FS pockets
become even larger in size than the LDA predications \cite{liebsch}.

The focus of the present work is to explain how strong correlations
drive orbital polarization and the band narrowing observed in ARPES.
We adopt a multi-orbital
Hubbard model description where the noninteracting part is determined by
fitting the LDA band structure. The interacting part contains both
the intra ($U$) and the inter-orbital ($U^\prime$) local Coulomb repulsion
as well as the Hund's rule coupling $J_H$.
First, a basis independent Hartree-Fock (HF) calculation is performed which is
in essence a LDA+U calculation in the paramagnetic phase. We find that
for $U^\prime$ much less than U, multi-orbital occupation is favored in order
to reduce the cost of double occupation. As a result, the HF self-energy
renormalizes the atomic level spacing in such a way that the size of the
FS pockets associated with the $e_g^\prime$ band grows. This trend is however
reversed when $U^\prime$ grows and becomes
comparable to U. In the HF theory, the size of the $e_g^\prime$ FS
pockets begins to shrink for $U^\prime/U > 3/5$. To correctly capture
the physics of strong correlation for large $U$ and $U^\prime$, we generalize
the Gutzwiller approximation to the case of multi-orbitals.
We find that in the strong-coupling regime, orbital polarization is
tied to Pauli-blocking, i.e. the orbital occupation dependence of the
Gutzwiller band renormalization factors. We obtain both band narrowing
and the disappearance of the FS pockets in good agreement with the
ARPES experiments.

We start with the multi-orbital tight-binding model on a two-dimensional
triangular lattice,
\begin{equation}
H_0=-\sum_{ij,\sigma}\sum_{\alpha \beta}t_{ij,\alpha \beta}
d^{\dagger}_{i\alpha \sigma}d_{j\beta \sigma}
-{\Delta \over 3}\sum_{i,\sigma}\sum_{\alpha \neq \beta}
d^{\dagger}_{i\alpha \sigma}d_{i\beta \sigma},
\label{h-tb}
\end{equation}
where the operator $d^{\dagger}_{i\alpha\sigma}$ creates an electron in the
$\alpha$ orbital with spin $\sigma$ on the Co site and
$t_{ij,\alpha\beta}$ is the hopping integral between the $\alpha$
orbital on site $i$ and the $\beta$ orbital on site $j$.
The relevant valence bands near the FS consist of
the Co $t_{2g}=\{d_{xy},d_{yz},d_{zx}\}$ orbitals and have an electron
occupancy of $5+x$. The $\Delta$ in Eq.~(\ref{h-tb}) describes
the crystal field due to trigonal distortion that splits the $t_{2g}$
complex into a lower $a_{1g}$ singlet and a higher $e'_g$ doublet,
where $a_{1g}=(d_{xy}+d_{yz}+d_{zx})/\sqrt{3}$,
and $e'_g=\{(d_{zx}-d_{yz})/\sqrt{2}$, $(2d_{xy}-d_{yz}-d_{zx})/\sqrt{6}\}$.
For convenience, we will work in the hole-picture via a particle-hole
transformation $d \rightarrow \tilde d^\dagger$, in which the band
filling of holes is $1-x$. The structure of
the tight-binding Hamiltonian in k-space on the triangular lattice is
\begin{equation}
H_0=\sum_{k,\sigma,\alpha\beta}
{\bf K}^d_{\alpha\beta}(k)
\tilde d^{\dagger}_{k\alpha\sigma}\tilde d_{k\beta\sigma}
+{\Delta\over 3}\sum_{k,\sigma,\alpha\neq\beta}
\tilde d^{\dagger}_{k\alpha\sigma}\tilde d_{k\beta\sigma}.
\label{hk-tb}
\end{equation}
The hopping matrix $\bf K$ in the $t_{2g}$ basis is given by
\begin{equation}
{\bf K}^d(k)=\left ( \begin{array}{clcr}
\varepsilon(t,1,2,3) &\varepsilon(t',3,1,2) &\varepsilon(t',2,3,1)\\
\varepsilon(t',3,1,2) &\varepsilon(t,2,3,1) &\varepsilon(t',1,2,3)\\
\varepsilon(t',2,3,1) &\varepsilon(t',1,2,3) &\varepsilon(t,3,1,2)
\end{array} \right),
\label{htk}
\end{equation}
with $(1,2,3)=(k_1,k_2,k_3)$, $k_1=\sqrt{3}k_x/2-k_y/2$,
$k_2=k_y$, $k_3=-k_1-k_2$, and
$\varepsilon(t,1,2,3)=2t_1\cos{k_1}+2t_2(\cos{k_2}+\cos{k_3})
+2t_3\cos{(k_2-k_3)}+2t_4[\cos{(k_3-k_1)}+\cos{(k_1-k_2)}]
+2t_5\cos{(2k_1)}+2t_6[\cos{(2k_2)}+\cos{(2k_3)}]+\cdots$.
The $(t,t')$ denote the (intra,inter)-orbital hopping.

\begin{figure}
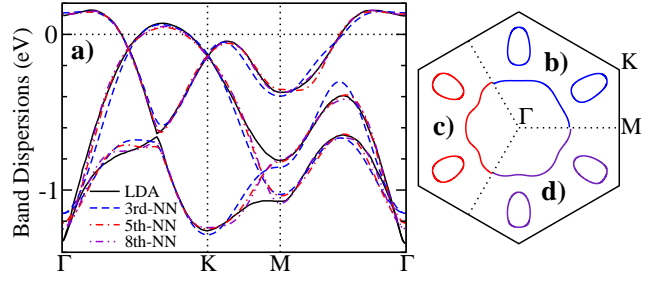

\begin{center}
\fig{3.3in}{fig1.eps}
\vskip-2.4mm
\caption{Tight-binding fits to the LDA band structure at $x=1/3$.
(a) The fitted band dispersions with up to 3rd, 5th and 8th NN hopping.
The corresponding Fermi surfaces are plotted in (b), (c), and (d)
respectively.}
\label{fig1}
\vskip-9mm
\end{center}
\end{figure}

Fig.{\ref{fig1}} shows the fitting of the tight-binding dispersions
obtained by diagonalizing Eq.~(\ref{hk-tb}) to the LDA band structure
at $x$=1/3 \cite{pickett}. We note that the fit with up to
third-nearest-neighbor (NN) hopping or more describes the LDA bands
quite well. On the other hand, the
tight-binding model cannot reproduce completely the LDA dispersions
even with up to eighth-NN hopping.
The discrepancy is most pronounced along the M-K
direction where the two $e'_g$ bands cross in the tight-binding fit
(Fig.~1a). Similar disagreement can be traced back to the previous
tight-binding fits \cite{yanase,liebsch,johannes}.
We believe the difficulty arises from the hopping path via the O 2$s$ and 2$p$
orbitals. Nevertheless, the tight-binding model works
very well at low energies near the Fermi level.
The FS consists of a cylindrical sheet around the
$\Gamma$-point and six hole pockets near the K-points as shown in Fig.~1b-d.
The central FS has a dominant $a_{1g}$ character while the six FS
pockets are mainly of the $e_g^\prime$ character.
The hopping integrals obtained from the fit with up to third-NN are
$t$=(-44.6, -9.0, 36.2, 5.9, 57.9, 36.7)meV and
$t'$=(-157.8, -30.2, 37.1, 9.2, -11.9, -21.0)meV.
The crystal-field splitting $\Delta$ is chosen to be 0.01eV.
In the rest of the paper, we use these parameters for $H_0$.
Our results are insensitive to these values provided that
they provide a good fit of the LDA band structure near the FS.
Note that although holes are evenly distributed among the three
$t_{2g}$ orbitals, in the $a_{1g}$ and $e_{g}^\prime$ basis
(hereafter referred to as the $\{a\}$ basis), the hole occupations
are 0.123 ($e_g^\prime$) and 0.421 ($a_{1g}$) respectively.
Despite its higher orbital energy, the $a_{1g}$ hole
orbital has a higher occupation due to its larger bandwidth.

The correlation effects are described by
the multi-orbital Hubbard model $H=H_0+H_I$, where
$H_0$ is the tight-binding Hamiltonian in Eq.(\ref{hk-tb}),
$H_I$ represents the local Coulomb repulsion $U$ (intra-orbital) and
$U'$ (inter-orbital) and Hund's rule coupling $J_H$.
For $t_{2g}$ orbitals, $H_I$ has been shown to take the
form \cite{castellani}
\begin{eqnarray}
H_I&=&U\sum_{i,\alpha}{\hat n}_{i\alpha\uparrow}{\hat n}_{i\alpha\downarrow}
+(U'-{1\over 2}J_H)\sum_{i,\alpha>\beta}{\hat n}_{i\alpha}{\hat n}_{i\beta}
\label{h-int} \\
&-&J_H\sum_{i,\alpha\neq\beta}{\bf S}_{i\alpha}\cdot {\bf S}_{i\beta}
+J_H\sum_{i,\alpha\neq\beta}a^{\dagger}_{i\alpha\uparrow}
a^{\dagger}_{i\alpha\downarrow}a_{i\beta\downarrow}a_{i\beta\uparrow}.
\nonumber
\end{eqnarray}
with $U^\prime = U - 2J_H$.
Here ${\hat n}_{i\alpha}$ and ${\bf S}_{i\alpha}$ are the density and
the spin operators in the $\{a\}$ basis where
the tight-binding part $H_0$ is
\begin{equation}
H_0=\sum_{k,\sigma}\sum_{\alpha\beta}
{\bf K}^a_{\alpha\beta}(k)
a^{\dagger}_{k\alpha\sigma}a_{k\beta\sigma}
+\sum_{k,\alpha,\sigma}\Delta_{\alpha}
a^{\dagger}_{k\alpha\sigma}a_{k\alpha\sigma}.
\label{hk-tba}
\end{equation}
Here $\Delta_{\alpha}$ = $-\Delta/3$, $2\Delta/3$ for the $e'_g$ and $a_{1g}$
orbitals respectively. The hopping matrix
${\bf K}^a(k)={\bf O}^{T}{\bf K}^d(k){\bf O}$,
with ${\bf O}$ the orthogonal rotation from the $t_{2g}$ to
the $\{a\}$ basis. $H_I$ is identical in
these two bases. The hierarchy of the interaction
strength is $U>U'>J_H\geq 0$.
\begin{figure}
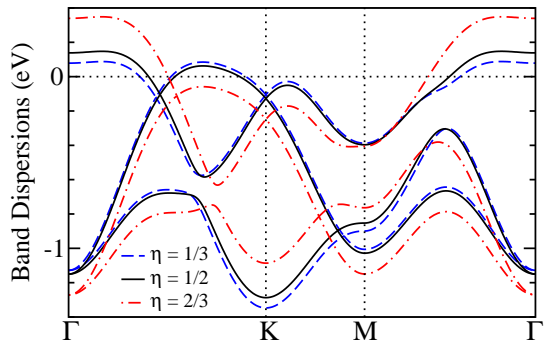

\begin{center}
\fig{2.8in}{fig2.eps}
\vskip-2.4mm
\caption{HF results for $U$ = 3.0\textit{eV} at $x=1/3$.
The band dispersions are shown for $\eta= 1/3, 1/2, 2/3$.
}
\label{fig2}
\vskip-9mm
\end{center}
\end{figure}

We first study the effects of interactions in the HF theory in the
orbital sector. In the paramagnetic phase, the interacting Hamiltonian
is given by,
\begin{eqnarray}
H_I^{HF} &=& \sum_{k,\sigma,\alpha}
({1\over 2}Un_{\alpha}+U_{\rm eff}^\prime
\sum_{\beta\neq\alpha}n_{\beta})
a^{\dagger}_{k\alpha\sigma}a_{k\alpha\sigma}\nonumber \\
&-&{U\over 4}\sum_{k,\alpha}n^2_{\alpha}
-{U_{\rm eff}^\prime\over 2}\sum_{k,\alpha\neq\beta}
n_{\alpha}n_{\beta}
\label{hartree} \\
&+&(U-2U_{\rm eff}^\prime)\sum_{k,\alpha\neq\beta}
\bigl[n_{\alpha\beta}
a^{\dagger}_{k\alpha\sigma}a_{k\beta\sigma}
-{n^2_{\alpha\beta}\over2}\bigr],
\nonumber
\end{eqnarray}
where $n_{\alpha\beta}
=(1/N_s)\sum_{k,\sigma}n_{\alpha\beta}^\sigma(k)$,
$n_{\alpha\beta}^\sigma(k)=\langle a^{\dagger}_{k\alpha\sigma}a_{k\beta\sigma}
\rangle$, $n_{\alpha}=n_{\alpha\alpha}$, and $U_{\rm
eff}^\prime=U^\prime-J_H/2$.
In essence, this HF analysis, also discussed in Ref.~\cite{liebsch},
is equivalent to the LDA+U theory \cite{LDA+U}.
Since we are interested in the orbital dependent corrections, we have not
displayed in Eq.~(\ref{hartree}) the double-counting term which corrects
for the energy already included in the LDA, because it depends only on the
total density.

Note that the HF theory is basis independent. It is thus convenient to
stay in the $\{a\}$ basis where the local density matrix, and thus the
HF self-energies are diagonal in orbital space, i.e.
$n_{\alpha\ne\beta}=0$.
Then the average energy has contributions only from the first two
terms in Eq.~(\ref{hartree}). Recall that most of the holes reside in
the $a_{1g}$ orbitals. If $U^\prime_{\rm eff} < U/2$, it is
favorable to increase the population of the $e_g^\prime$ orbitals to
take advantage of the smaller inter-orbital repulsion $U^\prime_{\rm
eff}$. On the other hand for $U^\prime_{\rm eff} > U/2$, the
tendency is to empty the $e_g^\prime$ orbitals in favor of $a_{1g}$
occupation. The crucial factor of $1/2$ in front of $U$ comes
from the fact that due to exchange, intra-orbital repulsion operates
only between holes with opposite spin, whereas both spins contribute
to $U_{\rm eff}^\prime$.

We proceed to calculate the HF self-energy
in terms of the average ${\bar n}=(n_{a_{1g}}+2n_{e_g^\prime})/3=(1-x)/3$
and the difference $\delta=(n_{a_{1g}}-n_{e'_g})/3$ between the
hole occupation of the $a_{1g}$ and $e_g^\prime$ orbitals,
\begin{eqnarray}
\Sigma^{\text{HF}}_{e'_g} &=&{1\over2}{\bar n}U(1+4\eta)+\delta U
(\eta-1/2)
\\
\Sigma^{\text{HF}}_{a_{1g}}&=&{1\over2}{\bar n}U(1+4\eta)-2\delta U
(\eta-1/2)
\end{eqnarray}
where $\eta=U_{\rm eff}^\prime/U$ is the relative strength of the
inter-orbital interaction.
The interaction effect in the paramagnetic HF theory
is to simply shift the atomic levels
by $\Sigma^{\text{HF}}_{e'_g}$ and $\Sigma^{\text{HF}}_{a_{1g}}$
respectively, resulting in a renormalization of the atomic level spacing
$\Delta^\prime=-3\delta U(\eta-1/2)$.
As expected, the direction
of the charge transfer depends on the ratio $\eta$.
Since the majority of the holes resides in
the $a_{1g}$ orbital in the noninteracting
limit, $\delta>0$.
Thus, for $\eta<1/2$, the level splitting renormalizes upward,
$\Delta^\prime >0$, and interactions induce a transfer of carriers
from the $a_{1g}$ to the $e_{g}^\prime$ orbital. The self-consistent
HF results are shown in Fig.~2 at $x=1/3$ for $U=3$eV which is close
to the value ($3.7$ eV) estimated from the LDA
\cite{johannes}. For $\eta=1/3$, the size of the hole pockets
indeed becomes larger than that of the noninteracting/LDA ones.
At $\eta=1/2$, the
self-energy corrections are equal among the orbitals and
the noninteracting (LDA) band dispersions are unchanged as shown in Fig.~2.
When $\eta>1/2$, i.e. for $U^\prime/U>3/5$ or $J_H/U<1/5$ which is
reasonable for the cobaltates \cite{PHZhang}, the level splitting
renormalizes downward, $\Delta^\prime <0$,
triggering a transfer of holes from the $e'_g$ to the $a_{1g}$ orbital.
The six FS pockets continues to shrink as the $e_g^\prime$ band
sinks with increasing $\eta$ and disappears beyond a critical ratio $\eta_c$,
as shown in Fig.~2 for $\eta=2/3$.
We find that $\eta_c(U=3.0{\rm eV})\simeq 5/8$.
\begin{figure}
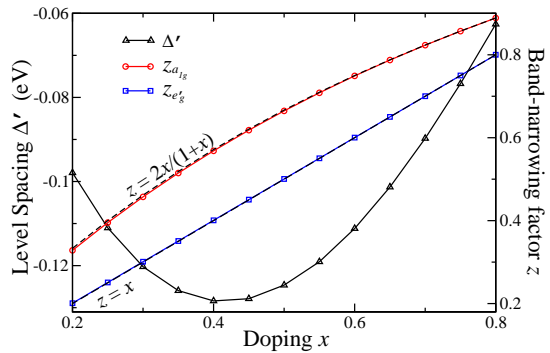

\begin{center}
\fig{2.8in}{fig3.eps}
\vskip-2.4mm
\caption{The doping $x$ dependence of the level spacing and bandwidth
renormalization for the $a_{1g}$ and $e_g^\prime$ orbitals.}
\label{fig3}
\vskip-9mm
\end{center}
\end{figure}

\begin{figure}
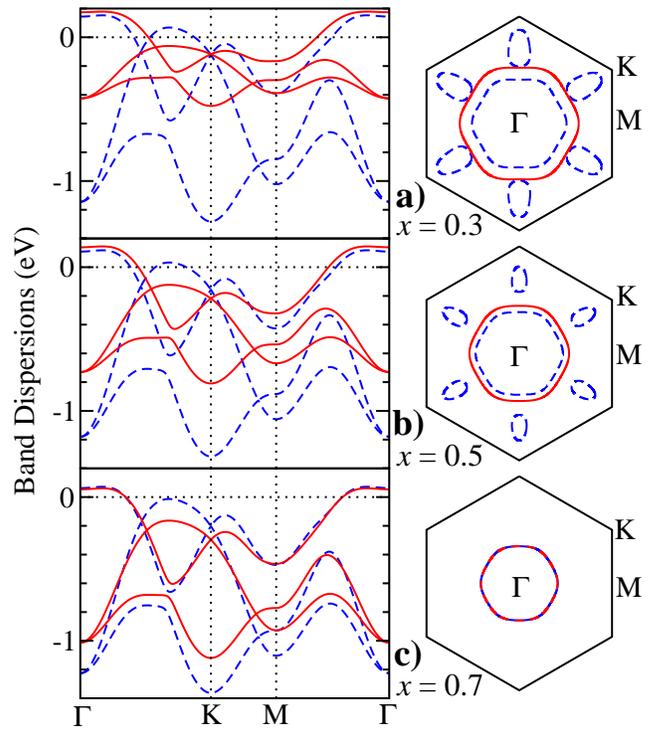

\begin{center}
\fig{3.3in}{fig4.eps}
\vskip-2.4mm
\caption{The band dispersions and the Fermi surfaces in the strong
coupling limit (red solid lines) for doping $x$ = 0.3, 0.5 and 0.7.
The noninteracting dispersions (blue dashed lines) are also plotted
for comparison.}
\label{fig4}
\vskip-9mm
\end{center}
\end{figure}

The HF analysis shows that the disappearance of the six
FS pockets near the K-points is the physics of large $U$ and $U'$
compared with $J_H$.
In this case, the HF theory itself becomes unreliable. Moreover, the
localization tendency leading to the bandwidth reduction cannot be
captured by the HF theory. It is therefore instructive to study the
problem in the strong-coupling limit by projecting out the states of
double-occupation prohibited by the large on-site Coulomb repulsions.
This can be achieved by the Gutzwiller projection
$\vert \Psi(\{n_\alpha\})\rangle
=P_G \vert \Psi_0(\{n_\alpha\})\rangle$, where $P_G$ is the
projection operator that operates on
the noninteracting state $\vert\Psi_0(\{n_\alpha\})\rangle$
in a given orbital occupation scheme $\{n_\alpha\}$. It removes the
double-occupancy by electrons from both the same and different orbitals.
This variational procedure is most conveniently implemented in
the Gutzwiller approximation where the effect of projection is
taken into account by the statistical weighting factor multiplying
the quantum coherent state \cite{fczhang}. Specifically,
we approximate the hopping term by
\begin{equation}
\langle\Psi\vert a_{i\alpha\sigma}^\dagger
a_{j\beta\sigma}\vert \Psi\rangle
=g_t^{\alpha\beta} \langle \Psi_0\vert
a_{i\alpha\sigma}^\dagger a_{j\beta\sigma}
\vert \Psi_0\rangle,
\label{gwt}
\end{equation}
where the Gutzwiller renormalization factor $g_t$ is given by the
ratio of the probabilities in the hopping process in the
projected $\vert\Psi\rangle$ and
the unprojected $\vert\Psi_0\rangle$. We find,
\begin{equation}
g_t^{\alpha\beta}={x\over\sqrt{(1-n_{i\alpha\sigma})(1-n_{j\beta\sigma})}}.
\label{gt}
\end{equation}
It is important to note that in a multi-orbital system $g_t^{\alpha\beta}$
depends on the occupation of the orbitals connected by the hopping
integral as seen in the denominator in Eq.~(\ref{gt}). The latter
originates from the Pauli principle.
It compensates for the effects of
``Pauli-blocking'' of double-occupation by electrons in the same
quantum states which already operate in the free fermion term on the RHS of
Eq.~(\ref{gwt}), while the numerator $x$ in Eq.~(\ref{gt}) 
describes the ``Coulomb-blocking''
due to the large on-site $U$ and $U^\prime$. It turns out that
the denominator is crucial for carrier transfer and orbital-polarization in the
strong-coupling limit. In the uniform paramagnetic phase, $g_t^{\alpha\beta}
=2x/\sqrt{(2-n_\alpha)(2-n_\beta)}$. The orbital occupations are variational
parameters determined by minimizing the ground state energy
of the Hamiltonian,
\begin{eqnarray}
H_{GW}&=&\sum_{k,\sigma,\alpha\beta}
g_t^{\alpha\beta}
{\bf K}^a_{\alpha\beta}(k)
a^{\dagger}_{k\alpha\sigma}a_{k\beta\sigma}
+\sum_{i,\alpha,\sigma}\Delta_{\alpha}
a^{\dagger}_{i\alpha\sigma}a_{i\alpha\sigma}
\nonumber \\
&+& \sum_{i,\alpha}\varepsilon_{i\alpha}
(\sum_\sigma a_{i\alpha\sigma}^\dagger a_{i\alpha\sigma}
-n_{i\alpha}),
\label{hgw}
\end{eqnarray}
where $\varepsilon_{i\alpha}$ are the Lagrange multipliers enforcing
the occupation $n_{i\alpha}=\sum_\sigma\langle
a_{i\sigma\alpha}^\dagger a_{i\sigma\alpha}
\rangle$. They are determined by the self-consistency equation
\begin{equation}
\varepsilon_{i\alpha}=\varepsilon_\alpha
={1\over 2-n_\alpha}{1\over N_s}\sum_{k,\beta,\sigma}
g_t^{\alpha\beta} {\bf K}^a_{\alpha\beta}
\langle a^{\dagger}_{k\alpha\sigma}a_{k\beta\sigma}\rangle.
\label{epsilon}
\end{equation}
The RHS of this equation is the derivative of the kinetic energy of
band $\alpha$ with respect to $n_\alpha$, and can be understood by
the following argument. The transfer of a hole from band $\alpha$ to
$\beta$ causes the respective bandwidths to decrease and increase by
${\cal O} (1/N_s)$. However, the kinetic energies of the occupied
states in each band are changed by order unity, and this energy
difference must be reflected in the equilibrium condition by an
energy shift $\varepsilon_\alpha$. Thus in contrast to HF theory,
there is no energy cost proportional to $U$. Instead,
both the band-narrowing and the renormalization of the level spacing 
$\Delta^\prime=\varepsilon_{a_{1g}}-\varepsilon_{e_g^\prime}$
contribute to the redistribution of holes among the orbitals.
In Fig.~3, we show the self-consistently determined band-narrowing factor
$z_\alpha=g_t^{\alpha\alpha}$ and the renormalized level spacing.
For orbitals with a larger hole occupation, the bandwidth reduction
is smaller and the renormalized band energy is lower, resulting
in the transfer of more holes into these bands. The combined
effects cause the holes to move out of the $e_g^\prime$ band into
the $a_{1g}$ band.
The calculated band dispersions and the FS topology at
$x=0.3, 0.5, 0.7$ are shown in Fig.{\ref{fig4}} in the strong
coupling theory. The corresponding non-interacting case is also shown
for comparison. The six hole pockets are completely absent at all
levels of sodium doping due to strong correlation, leaving a single hexagonal
Fermi surface centered around the $\Gamma$ point satisfying the
Luttinger theorem. This, as well as the band-narrowing due to strong
Coulomb repulsion, is in very good agreement with the photoemission
experiments \cite{hbyang2,hasan2}.

In conclusion, we have shown that strong correlation plays
an important role in understanding the electronic structure of Na$_x$CoO$_2$.
It pushes the $e_g^\prime$ band below
the Fermi level, leading to an orbital polarized state with a single
hole-like FS. The absence of the small FS pockets, which
would have contributed significantly to the density of states in band
structure calculations, further suggests that the large mass enhancement
observed in the specific heat measurement \cite{chou}
is due to strong correlation.

This work is supported by DOE grants DE-FG02-99ER45747 and DE-FG02-03ER46076,
ACS grant 39498-AC5M, and NSF grants DMR-0072205 and DMR-0201069.

\end{document}